\documentclass[aps,preprint,showpacs,superscriptaddress,groupedaddress]{revtex4}  
\usepackage[dvips]{graphicx}
\usepackage{dcolumn}
\usepackage{hyperref}   
\usepackage{bm}        
\usepackage{amssymb}   
\usepackage[utf8]{inputenc}
\usepackage{mathrsfs}
\usepackage{amsmath}
\usepackage{hyperref}

\begin{document}

\widetext

\title{Kink interactions in the (1+1)-dimensional $\varphi^6$ model}
\author{Vakhid A. Gani}
\email{vagani@mephi.ru}
\affiliation{National Research Nuclear University MEPhI (Moscow Engineering Physics Institute), 115409 Moscow, Russia}
\affiliation{National Research Center Kurchatov Institute, Institute for Theoretical and Experimental Physics, 117218 Moscow, Russia}
\author{Alexander E. Kudryavtsev}
\email{kudryavt@itep.ru}
\affiliation{National Research Center Kurchatov Institute, Institute for Theoretical and Experimental Physics, 117218 Moscow, Russia}
\author{Mariya A. Lizunova}
\email{mary.lizunova@gmail.com}
\affiliation{National Research Nuclear University MEPhI (Moscow Engineering Physics Institute), 115409 Moscow, Russia}
\affiliation{National Research Center Kurchatov Institute, Institute for Theoretical and Experimental Physics, 117218 Moscow, Russia}

\vskip 1cm

\begin{abstract}
We study kink scattering processes in the (1+1)-dimensional $\varphi^6$ model in the framework of the collective coordinate approximation. We find critical values of the initial velocities of the colliding kinks. These critical velocities distinguish different regimes of collisions. The exact equation of motion for the $\varphi^6$ model is also solved numerically with the same initial conditions. We discuss advantages and disadvantages of the collective coordinate approximation, and also outline its applicability limits. Resonance phenomena and the so-called escape windows are also observed in the kink collisions.
\end{abstract}

\maketitle

\section{\label{sec:level1} Introduction}

Topological defects are of growing interest in physics. They arise in a great amount of models from classical and quantum field theory to condensed matter \cite{vilenkin01,aek01,bazeia01}. In this context, one-dimensional topological defects play a special role: on the one hand, many physical phenomena involve the formation of effectively one-dimensional topological structures, for example, a three- or two-dimensional domain wall in the direction perpendicular to it presents a one-dimensional topological field configuration (kink) interpolating two different vacua of the model. On the other hand, the case of (1+1) dimensions can be more easily investigated, hence the use of (1+1)-dimensional systems as simplified models. Note that it is an actively developing area, with many important results obtained recently: the topological defect deformation procedure \cite{bazeia02,bazeia03}, the construction of a topological defect carrying $U(1)$ charge in a system with two scalar fields -- one real and one complex \cite{gani02,gani03}, $Q$ balls in scalar theories with $U(1)$ symmetry \cite{mai01,nugaev}, and many others. There are interesting results in scalar systems with an interaction with a spinor field \cite{stoj,gani04,gani05}. Apart from that, supersymmetric models with more complex vacuum manifolds and several types of kinks connecting the different vacua have also been discussed \cite{gani01}. Special attention should be paid to configurations of strings or vortices \cite{vilenkin01,kibble01,kibble02}.  In particular, there are important results on quantum stabilization of strings \cite{weigel02,weigel03,weigel04} and strings/vortices/monopoles dynamics \cite{palvelev,dziarmaga,myers,sutcliffe,ward}.

The collective coordinate method has been successfully applied to study solitary wave interactions in various systems. The idea of the method is to treat an originally constant parameter of a kink (for example, its position) as a time-dependent variable that we will call a collective coordinate. Originally, this method was used to describe kink-kink interactions in the $\lambda\varphi^4$ theory \cite{aek02}. It was also applied to study collisions of domain walls in a supersymmetric model \cite{gani01}. The problem of the collision of parallel domain walls was solved in the collective coordinate approximation with the distance between domain walls being the single collective coordinate. For sufficiently slow collisions, the results of the collective coordinate approach agreed well with the exact numerical solution of the equations of motion. This work also showed that, depending on the initial velocity, there are two different regimes of the evolution of the system, and found the critical velocity that separates these regimes. In Ref.~\cite{baron01} the collective coordinate approximation was applied to the interaction of two solitons of the nonlinear Schrodinger equation. In this context, see also Refs.~\cite{hata01,hata02}, which develop a relativistic generalization of the collective coordinate method and perform the quantization of the rotational motion of the skyrmion. A general discussion of the collective coordinate method involving the discrete (vibrational) mode of a kink is given in the review \cite{aek01}.

In this work we study the kink scattering in the (1+1)-dimensional $\varphi^6$ model in the framework of the collective coordinate approximation. We consider different topologies and various initial velocities of the colliding kinks. We compare the results obtained in the framework of the collective coordinate approximation with those obtained by solving the field equations numerically at the same initial conditions. We analyze the differences and outline the scope of applicability of the collective coordinate approximation. Solving the exact equations of motion numerically, we observe the so-called escape windows \cite{dorey01,weigel}; note that this feature of the kink scattering cannot be described by the collective coordinate approximation with $1$ degree of freedom. Our analysis does not confirm the discrepancy between the results of the collective coordinate approximation and a numerical solution of the exact equation of motion, reported recently for one of the topological sectors of the model [the type $(-1,0,1)$, see below in Sec.~\ref{sec:level5}] by the author of Ref.~\cite{goatham}.

Our paper is organized as follows. In Sec.~\ref{sec:level2} we provide some general facts about topological localized solutions with finite energy in different models in (1+1) dimensions. Section~\ref{sec:level3} is devoted to the description of the properties of the $\varphi^6$ model and its static topological solutions (kinks). In Sec.~\ref{sec:level4} we give the details of the collective coordinate method and formulate the problem of a collision of two kinks. In Sec.~\ref{sec:level5}, we present our main results, along with some technical details of our calculations, and compare them with the results of other authors. We conclude with a discussion of the results and prospects for future research.

\section{\label{sec:level2} Static solutions in (1+1) dimensions}

We consider a real scalar field $\varphi(t,x)$ in $(1 +1)$ dimensions, with its dynamics determined by the Lagrangian
\begin{equation}
	\mathscr{L}=\frac{1}{2} \left( \frac{\partial\varphi}{\partial t} \right)^2-\frac{1}{2} \left( \frac{\partial\varphi}{\partial x} \right) ^2-U(\varphi),		
	\label{eq:largang}
\end{equation}
where $U(\varphi)$ is a potential, defining the self-interaction of the field $\varphi$. The Lagrangian~(\ref{eq:largang}) yields the equation of motion for $\varphi(t,x)$:
\begin{equation}
	\frac{\partial^2\varphi}{\partial t^2}-\frac{\partial^2\varphi}{\partial x^2}+\frac{dU}{d\varphi}=0.
	\label{eq:eom}
\end{equation}
The vacua of the model are defined by the minimal points of $U(\varphi)$: $\varphi_0^{(1)}$, $\varphi_0^{(2)}$, etc. Further we consider a model with a non-negative potential $U(\varphi)$ having two or more degenerate minima with $U_{\scriptsize\mbox{min}}=0$.

If we are interested in static solutions $\varphi=\varphi(x)$, then Eq.~(\ref{eq:eom}) becomes
\begin{equation}
	\frac{d^2\varphi}{dx^2}=\frac{dU}{d\varphi}.
	\label{eq:steom}
\end{equation}
The energy functional for the Lagrangian~(\ref{eq:largang}) is
\begin{equation}
	E[\varphi]=\int_{-\infty}^{\infty}\left[\frac{1}{2} \left( \frac{\partial\varphi}{\partial t} \right)^2+\frac{1}{2} \left( \frac{\partial\varphi}{\partial x} \right) ^2+U(\varphi)\right]dx.
\end{equation}
For static fields $E[\varphi]$ takes the form
\begin{equation}
	E[\varphi]=\int_{-\infty}^{\infty}\left[\frac{1}{2} \left( \frac{d\varphi}{d x} \right) ^2+U(\varphi)\right]dx.
	\label{eq:stenerg}
\end{equation}
In order for the integral in~(\ref{eq:stenerg}) to be convergent, i.e.,\ for the configuration energy to be finite, it is necessary that

\begin{equation}
	\lim_{x \to -\infty} \varphi (x) = \varphi_0^{{\scriptsize \mbox{(i)}}}
	\label{eq:doubletwo}
\end{equation}
and
\begin{equation}
	\lim_{x \to +\infty} \varphi (x) = \varphi_0^{{\scriptsize \mbox{(j)}}}\\.
	\label{eq:two}
\end{equation}
If (\ref{eq:doubletwo}) and (\ref{eq:two}) hold, then both terms in square brackets in~(\ref{eq:stenerg}) fall off at $x\to\pm\infty$ and the integral can be convergent.

From Eq.~(\ref{eq:steom}), one can easily obtain a first-order differential equation of motion
\[
\displaystyle\frac{1}{2}\left(\frac{d\varphi}{dx}\right)^2=U(\varphi),
\]
or
\begin{equation}
	\frac{d\varphi}{dx}=\pm\sqrt{2U}.
	\label{eq:five}
\end{equation}
If there are two or more degenerate minima of the potential $U(\varphi)$, the set of static solutions with finite energy splits into disjoint classes (topological sectors) according to their asymptotic behavior at $x\to\pm\infty$. Solutions with $\varphi(+\infty)\neq\varphi(-\infty)$ are called topological, while those with $\varphi(+\infty)=\varphi(-\infty)$ -- nontopological. A configuration belonging to one topological sector can not be transformed into a configuration belonging to another topological sector through a continuous deformation, i.e.,\ via a sequence of configurations with a finite energy. For more details see, e.g., Refs.~\cite{aek01,bazeia01}.

One can introduce a conserved topological current, for example,
\[
	j_{{\scriptsize \mbox{top}}}^\mu=\frac{1}{2}\varepsilon^{\mu\nu}\partial_\nu\varphi.
\]
The corresponding conserved topological charge is determined only by the asymptotics of the field $\varphi(x)$ and does not depend on its behavior at finite $x$:
\begin{equation}
	Q_{{\scriptsize \mbox{top}}}=\int_{-\infty}^{\infty}j_{{\scriptsize \mbox{top}}}^0dx=\frac{1}{2} \left[ \varphi(+\infty)-\varphi(-\infty) \right].
\end{equation}
Configurations with different topological charges necessarily belong to different topological sectors, however, configurations belonging to different topological sectors may have the same topological charge.

As already mentioned, the function $U(\varphi)$ is considered to be non-negative. This allows one to introduce the following useful definition:
\begin{equation}
	U(\varphi)=\frac{1}{2}\left(\frac{dW}{d\varphi}\right)^2,
	\label{eq:dwdfi}
\end{equation}
where $W(\varphi)$ is a smooth (continuously differentiable) function of the field $\varphi$ called the superpotential. Using the representation (\ref{eq:dwdfi}), the energy (\ref{eq:stenerg}) can be written as~\cite{bazeia01}
\[
	E=E_{\scriptsize \mbox{BPS}}+\frac{1}{2}\int_{-\infty}^{\infty}\left(\frac{d\varphi}{dx}\pm\frac{dW}{d\varphi}\right)^2dx
\]
with
\[
E_{\scriptsize \mbox{BPS}}=|W[\varphi(+\infty)]-W[\varphi(-\infty)]|,
\]
where the subscript ``BPS'' stands for Bogomolny, Prasad, Sommerfield \cite{bps}.
Thus the static configuration that satisfies the equation
\begin{equation}
	\displaystyle\frac{d\varphi}{dx}=\pm\frac{dW}{d\varphi}
	\label{eq:doublefive}
\end{equation}
has the minimal energy $E=E_{\scriptsize \mbox{BPS}}$ among all the configurations in a given topological sector. The solutions that satisfy Eq.~(\ref{eq:doublefive}) are called BPS-saturated (or simply BPS) configurations.

For a more detailed review of the properties of models with one scalar field in $(1+1)$ dimensions see, for example, \cite{vilenkin01,aek01,rajaraman,bazeia01}.

\section{\label{sec:level3} the $\varphi^{6}$ model}

\begin{figure}
	\includegraphics[scale=0.8]{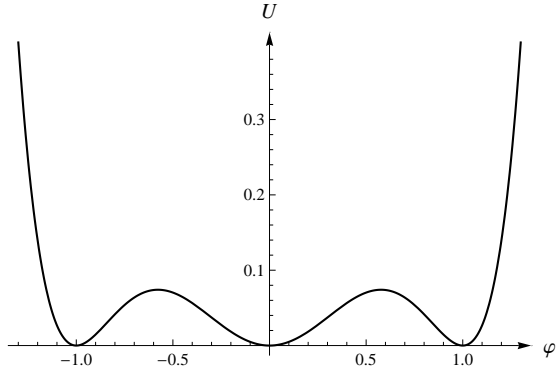}
	\caption{\label{fig:01} The potential of the $\varphi^{6}$ model.}
\end{figure}

Consider the $\varphi^{6}$ model with a real scalar field in $(1+1)$ dimensions \cite{lohe}, defined by the Lagrangian~(\ref{eq:largang}) with the potential
\begin{equation}
	U(\varphi)=\frac{1}{2}\varphi^2(1-\varphi^2)^2.
	\label{eq:potfi6}
\end{equation}
This potential has three minima -- vacua of the theory: $\varphi_0^{(1)}=-1$, $\varphi_0^{(2)}=0$ and $\varphi_0^{(3)}=+1$; see Fig.~\ref{fig:01}. According to Eq.~(\ref{eq:dwdfi}), the superpotential of this model is
\[
	W(\varphi)=\frac{\varphi^2}{2}-\frac{\varphi^4}{4}.
\]
Static topological solutions (kinks) can be easily found by solving the first-order differential equation of motion:
\begin{equation}
	\frac{d\varphi}{dx}=\pm\varphi(1-\varphi^2).
	\label{eq:fivefi6}
\end{equation}
Its solutions belonging to different topological sectors are
\begin{equation}
	\varphi(x-\tilde{x}_0)=\pm\sqrt{\frac{1}{2}(1\pm\tanh(x-\tilde{x}_0))}.
	\label{eq:unintkink}
\end{equation}
The constant $\tilde{x}_0$ is arbitrary due to the translational invariance of the system and is related to the position of the kink. Equation~(\ref{eq:unintkink}) can be rewritten in a slightly different form
\begin{equation}
	\displaystyle\varphi(x-x_0)=\frac{\pm 1}{\sqrt{1+3e^{\pm 2(x-x_0)}}}
	\label{eq:stkink1}
\end{equation}
that will be used below. The constants $x_0$ and $\tilde{x}_0$ are related by
\[
	x_0 = \tilde{x}_0 \pm \ln{\sqrt{3}}.
\]
For example, the kink
\[
	\varphi_{(0,1)}(x-x_0)=\frac{1}{\sqrt{1+3e^{ -2(x-x_0)}}}
\]
interpolates between the vacua $\varphi_0^{(2)}=0$ and $\varphi_0^{(3)}=1$, i.e.,\ it belongs to the topological sector $(0,1)$, while the kink
\[
	\varphi_{(-1,0)}(x-x_0)=-\frac{1}{\sqrt{1+3e^{2(x-x_0)}}}
\]
represents the sector $(-1,0)$, and so on. The mass of each kink is $\displaystyle\frac{1}{4}$. A moving kink can be obtained by the Lorentz boost.

\section{\label{sec:level4} Collective coordinate approximation}

The idea of the method as applied to (1+1)-dimensional systems is the following. The initial field configuration $\varphi(x)$ is chosen in the form of two kinks $\varphi_{\scriptsize \mbox{k}}^{(1)}$ and $\varphi_{\scriptsize \mbox{k}}^{(2)}$ that are far apart, i.e.,\ are separated by a large distance much greater than the typical scale of the kink, e.g.,
\begin{equation}
	\varphi(x) = \varphi_{\scriptsize \mbox{k}}^{(1)}(x+a)+\varphi_{\scriptsize \mbox{k}}^{(2)}(x-a)+\mbox{const}.
	\label{eq:anzac}
\end{equation}
This configuration is not an exact solution of the equation of motion, however, at $a\gg 1$ the overlap between the kinks is exponentially small, hence (\ref{eq:anzac}) is exact up to exponentially small terms.

Further let us suppose $a$ to be a function of time, $a=a(t)$, thus allowing the kinks to move towards or away from each other, with $2 a(t)$ being the distance between the kinks. As a result we have a system with $1$ degree of freedom $a(t)$. The dynamics of the system is governed by the Lagrange function $L_{{\scriptsize \mbox{eff}}}(a,\dot{a})$ that can be derived by the substitution of Eq.~(\ref{eq:anzac}) in the Lagrangian~(\ref{eq:largang}) followed by the integration over $x$. We emphasize that one must take into account that $a=a(t)$ when calculating $\displaystyle\frac{\partial\varphi}{\partial t}$. The resulting Lagrange function has the following general form:
\begin{equation}
	L_{{\scriptsize \mbox{eff}}}(a,\dot{a})=\frac{1}{2}m(a)\dot{a}^2-V(a).
	\label{eq:efflagrang}
\end{equation}
The specific dependencies $m(a)$ and $V(a)$ are determined by the model under consideration and by the initial configuration~(\ref{eq:anzac}). Apart from the equations of motion following from $L_{{\scriptsize \mbox{eff}}}(a,\dot{a})$, one needs to specify the initial separation between the kinks $a(0)$ and the initial speed $|\dot{a}(0)|$ of the kink.

The dependence $a(t)$ can be obtained by solving the Cauchy problem for the Euler-Lagrange equation which, in this case, is a second-order ordinary differential equation:
\[
 \frac{d}{dt}\frac{\partial L_{{\scriptsize \mbox{eff}}}}{\partial\dot{a}}-\frac{\partial L_{{\scriptsize \mbox{eff}}}}{\partial a}=0.
\]
For the effective Lagrange function (\ref{eq:efflagrang}) we have
\begin{equation}
	m\ddot{a}+\frac{1}{2}\frac{dm}{da}\dot{a}^2+\frac{dV}{da}=0.
	\label{eq:eiler}
\end{equation}

The collective coordinate approximation ansatz (\ref{eq:anzac}) does not take into account the Lorentzian change of the shape of the moving kinks. The framework of the collective coordinate approximation thus neglects the relativistic effects (as well as the excitation of the kinks' internal degrees of freedom). This restricts the applicability of the method to small initial velocities. The consequences of these approximations will be discussed in detail below.

Note that relativistic effects have been taken into account in some studies that applied the collective coordinate approximation. For example, the authors of Ref.~\cite{okun} used the collective coordinate method to study the evolution of a spherically symmetric domain (a bubble) of a vacuum immersed in a different vacuum, in a model with a spontaneously broken symmetry. They used a relativistic Lagrangian in order to describe the dynamics of the domain wall; the speed of the wall could reach ultrarelativistic values during the collapse of the bubble.

\section{\label{sec:level5} Kink collisions in the $\varphi^6$ model}

We applied the collective coordinate method to study the collisions of the $\varphi^6$ kinks. We consider the following kink-kink collisions: $(-1,0)$ and $(0,1)$, $(0,-1)$ and $(-1,0)$, $(-1,0)$ and $(0,-1)$. For future convenience, we write out here all the kinks interpolating the different vacua of the model. We denote the topological sector of the configuration by $(\varphi(-\infty),\varphi(+\infty))$, so that
\begin{table}[h!]
\begin{center}
\begin{tabular}{cc}
$\varphi_{(-1,0)}(x)=-\displaystyle\frac{1}{\sqrt{1+3e^{2x}}}$, & $\varphi_{(1,0)}(x)=\displaystyle\frac{1}{\sqrt{1+3e^{2x}}}$, \\
\\
$\varphi_{(0,-1)}(x)=-\displaystyle\frac{1}{\sqrt{1+3e^{-2x}}}$, & $\varphi_{(0,1)}(x)=\displaystyle\frac{1}{\sqrt{1+3e^{-2x}}}$. \\
\end{tabular}
\end{center}
\end{table} \\
Note that
\[
\varphi_{(1,0)}(x)=-\varphi_{(-1,0)}(x)=\varphi_{(0,1)}(-x)=-\varphi_{(0,-1)}(-x).
\]
In addition, we use the notation where, for example,
\[
	\varphi_{(0,1,0)}(x)=\varphi_{(0,1)}(x+a)+\varphi_{(1,0)}(x-a)-1
\]
is called the configuration of the type $(0,1,0)$, etc.

In what follows, we use the superscripts ``(eff)'', ``(eom)'', and ``(mech)'' to distinguish between the values of the critical velocities obtained, respectively, from a numerical solution of Eq.~(\ref{eq:eiler}), from a numerical solution of the exact equation of motion~(\ref{eq:eom}), or from classical mechanics arguments within the collective coordinate approximation~(\ref{eq:efflagrang}).

\subsection{\label{sec:level6} Evolution of the configuration $(-1,0,1)$}

\begin{figure}
	\includegraphics[scale=0.8]{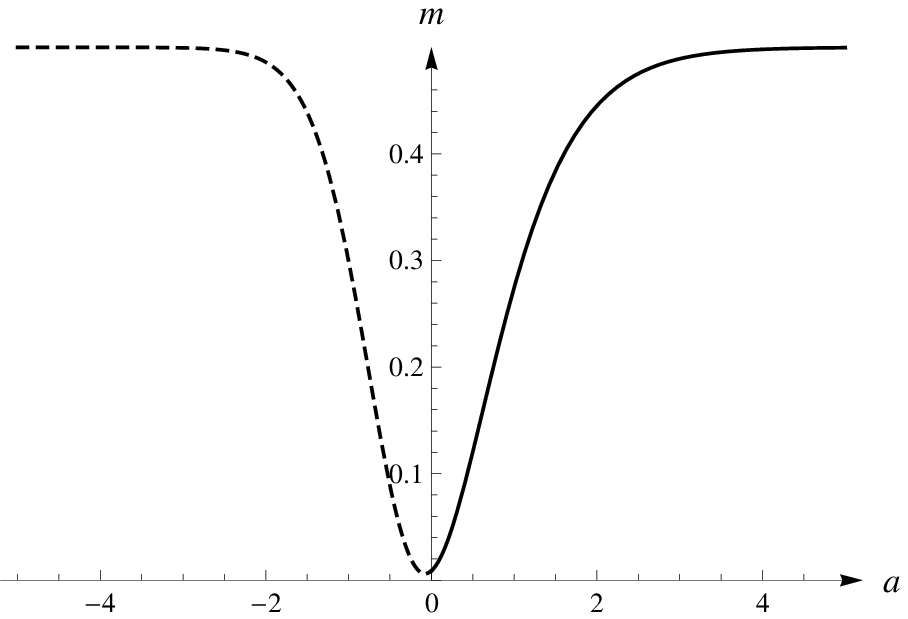}
	\caption{\label{fig:kink1-ma} The dependence $m(a)$ for the configuration~(\ref{eq:kink1}).}
\end{figure}

\begin{figure}
	\includegraphics[scale=0.8]{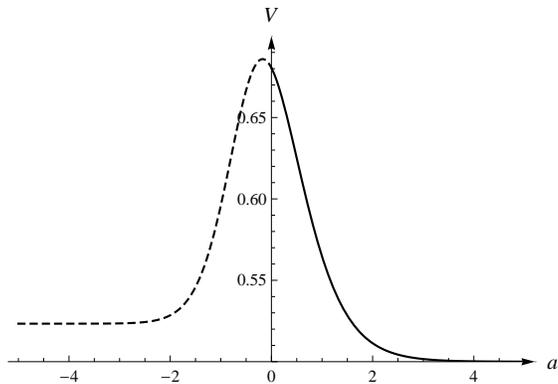}
	\caption{\label{fig:kink1-va} The dependence $V(a)$ for the configuration~(\ref{eq:kink1}).}
\end{figure}

The suitable collective coordinate approximation ansatz (\ref{eq:anzac}) is in this case
\begin{equation}
	\varphi_{(-1,0,1)}(x)=\varphi_{(-1,0)}(x+a)+\varphi_{(0,1)}(x-a).
	\label{eq:kink1}
\end{equation}
\begin{figure}
	\includegraphics[scale=0.8]{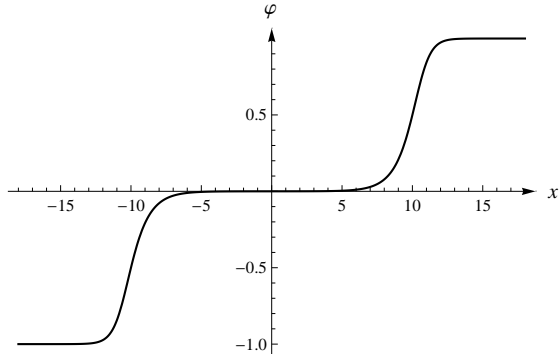}
	\caption{\label{fig:kink1} Ansatz~(\ref{eq:kink1}) at $a=10$.}
\end{figure}
The plot of $\varphi_{(-1,0,1)}(x)$ at $a=10$ is presented in Fig.~\ref{fig:kink1}. The effective Lagrange function (\ref{eq:efflagrang}) for $a(t)$ has the following $m(a)$ and $V(a)$:
\[
	m(a)=I_{-}(a),
\]
\[
	V(a)=\frac{1}{2}I_{+}(a)+\frac{1}{2}\int_{-\infty}^{\infty}\varphi_{(-1,0,1)}^2(x)(1-\varphi_{(-1,0,1)}^2(x))^2dx,
\]
where
\[
 I_{\pm}(a)=\frac{1}{2}\pm18e^{4a}\int_{-\infty}^{\infty}\frac{dx}{(1+3e^{-2(x-a)})^{3/2}(1+3e^{2(x+a)})^{3/2}}.
\]
The profiles of $m(a)$ and $V(a)$ are shown in Figs.~\ref{fig:kink1-ma} and \ref{fig:kink1-va}.

Evidently the potential $V(a)$ is repulsive. Therefore, we should expect to see an elastic reflection of the kinks, at least when the initial velocities are not too large. This is confirmed by our numerical analysis.

From the shape of $V(a)$, Fig.~\ref{fig:kink1-va}, we expect to find a critical value of the initial velocity $v_{{\scriptsize \mbox{cr}}}^{{\scriptsize \mbox{(eff)}}}$ of the colliding kinks. This critical velocity separates two different modes of the collision: at $v_{{\scriptsize \mbox{in}}}<v_{{\scriptsize \mbox{cr}}}^{{\scriptsize \mbox{(eff)}}}$ the elastic reflection should be observed, while at $v_{{\scriptsize \mbox{in}}}>v_{{\scriptsize \mbox{cr}}}^{{\scriptsize \mbox{(eff)}}}$ the kinks should pass through each other and escape to infinities with the final velocities $v_{\scriptsize \mbox{f}}<v_{\scriptsize \mbox{in}}$. Since the shapes of the $(-1,0)$ and $(0,1)$ kinks differ only slightly, in both cases ($v_{{\scriptsize \mbox{in}}}<v_{{\scriptsize \mbox{cr}}}^{{\scriptsize \mbox{(eff)}}}$ and $v_{{\scriptsize \mbox{in}}}>v_{{\scriptsize \mbox{cr}}}^{{\scriptsize \mbox{(eff)}}}$) we observe a collision of these two kinks which is followed by their escape. However, there is an essential difference between these two collision regimes: the kinks that are reflected elastically remain in their respective topological sectors, whereas the kinks that pass through each other exchange their topological sectors [of course, the type of the whole configuration, $(-1,0,1)$, does not change]. In other words at $v_{{\scriptsize \mbox{in}}}>v_{{\scriptsize \mbox{cr}}}^{{\scriptsize \mbox{(eff)}}}$, the kink $(-1,0)$ incident from the left interpolates between the vacua 0 and 1 after the collision, and analogously for the initial kink $(0,1)$ which switches to interpolate between the vacua $-1$ and 0. The difference between the shapes of the two initial kinks is crucial: it means that, after the collision, the kinks are no longer the exact solutions of their new topological sectors. Their masses are larger than the masses of these exact solutions, and this reflects itself in the fact that the potential $V(a)$ has different asymptotic values as $a\to\pm\infty$, $V(-\infty)>V(+\infty)$.

\begin{figure}
	\includegraphics[scale=0.8]{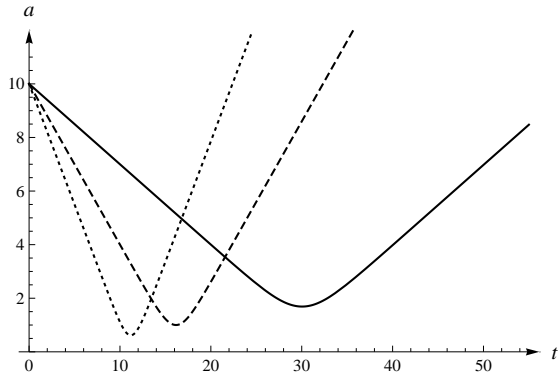}
	\caption{\label{fig:kink1-at} The half-distance $a$ between the kinks as a function of time $t$ for the configuration~(\ref{eq:kink1}) at $a(0)=10$ and different initial velocities: $v_{{\scriptsize \mbox{in}}}=0.3$ (solid curve), $v_{{\scriptsize \mbox{in}}}=0.6$ (dashed curve), $v_{{\scriptsize \mbox{in}}}=0.9$ (dotted curve). Collective coordinate approximation.}
\end{figure}

In Fig.~\ref{fig:kink1-at} we show $a(t)$ obtained numerically from Eq.~(\ref{eq:eiler}) for several initial velocities $v_{{\scriptsize \mbox{in}}}=|\dot{a}(0)|<v_{{\scriptsize \mbox{cr}}}^{{\scriptsize \mbox{(eff)}}}$. Since $v_{{\scriptsize \mbox{in}}}<v_{{\scriptsize \mbox{cr}}}^{{\scriptsize \mbox{(eff)}}}$, $a$ decreases to some $a_{\scriptsize \mbox{min}}>0$ and then begins to increase, as expected in the regime of reflection.

On the other hand, when $v_{\scriptsize \mbox{in}}>v_{{\scriptsize \mbox{cr}}}^{{\scriptsize \mbox{(eff)}}}$ then $a$ continues to decrease through zero and changes its sign. The change of the sign of $a$ means that the kink $(-1,0)$ has passed to the right and changed its sector to $(0,1)$, while the kink $(0,1)$ has passed to the left and changed its sector to $(-1,0)$.

Within the collective coordinate approximation, Eq.~(\ref{eq:eiler}) gives the value of the critical velocity $v_{\scriptsize \mbox{cr}}^{\scriptsize\mbox{(eff)}}>1$. The value of the critical speed is larger than the speed of light (recall $c=1$ in our units), which is obviously due to the Lagrange problem (\ref{eq:efflagrang}) not being Lorentz invariant [recall that the relativistic effects are not taken into account in our collective coordinate approximation ansatz (\ref{eq:anzac}), (\ref{eq:kink1})]. This extreme value clearly indicates that the issue of whether there exists a critical speed in this collision configuration cannot be answered by the collective coordinate method.

To check the accuracy of the collective coordinate approximation we solved the partial differential equation~(\ref{eq:eom}) numerically with the same initial conditions, i.e.,\ initial positions and velocities of the kinks. We note here that the Lorentz factor was taken into account for the numerical solution of Eq.~(\ref{eq:eom}).

\begin{figure}
	\includegraphics[scale=0.8]{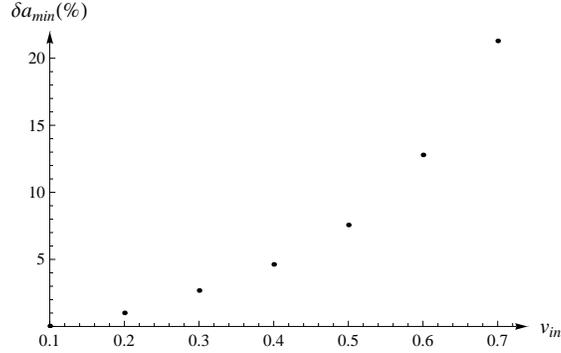}
	\caption{\label{fig:a_min-eom-col} The relative difference (\ref{eq:damin}) between the values of $a_{\scriptsize\mbox{min}}$, obtained by the two methods, as a function of the initial velocity $v_{\scriptsize\mbox{in}}$.}
\end{figure}

In Fig.~\ref{fig:a_min-eom-col}, we compare the values of $a_{\scriptsize\mbox{min}}(v_{\scriptsize\mbox{in}})$ resulting in the collective coordinate approximation with those obtained by the numerical solution of the equation of motion. This figure shows the relative difference between the values of $a_{\scriptsize\mbox{min}}$, obtained by the two methods, as a function of the initial velocity $v_{\scriptsize\mbox{in}}$:
\begin{equation}
	\delta a_{\scriptsize\mbox{min}}=\frac{a_{\scriptsize\mbox{min}}^{\scriptsize \mbox{(eff)}}-a_{\scriptsize\mbox{min}}^{\scriptsize \mbox{(eom)}}}{a_{\scriptsize\mbox{min}}^{\scriptsize \mbox{(eom)}}}\cdot 100\% .
	\label{eq:damin}
\end{equation}
Note that our results show significantly better agreement between the two methods than reported in Ref.~\cite{goatham}, where the collective coordinate approximation was found to overestimate the exact result by about 50\% at initial velocity $v_{\scriptsize \mbox{in}}=0.3$.

Solving Eq.~(\ref{eq:eom}) numerically, we could not find a critical velocity $v_{\scriptsize \mbox{cr}}^{\scriptsize\mbox{(eom)}}$ that would separate the reflection and transition scattering regimes. However, we conjecture this value to be very close to unity. The difficulty that we encountered in trying to distinguish between the reflection and transition regimes is apparently due to the fact that in the latter regime the two kinks that pass through each other and change their topological sectors quickly evolve into the respective exact solutions of their new sectors. The emerging configuration thus becomes indistinguishable from that corresponding to a reflection of the two kinks. See the next section for a more detailed discussion of the evolution of the kinks after they transit through each other and change the topological sectors, thus ceasing to be exact solutions of their sectors.

We note here that we could not observe the passage of the two colliding kinks through each other; hence, the conclusions about the details of this process are rather speculative. This is why in Figs.~\ref{fig:kink1-ma} and \ref{fig:kink1-va} the sections of the curves corresponding to negative values of $a$ are shown by dashed lines, so as to stress that the evolution of the ansatz in this region of the collective coordinate remains to be investigated.

\subsection{\label{sec:level7} Evolution of the configuration $(0,-1,0)$}

\begin{figure}
	\includegraphics[scale=0.8]{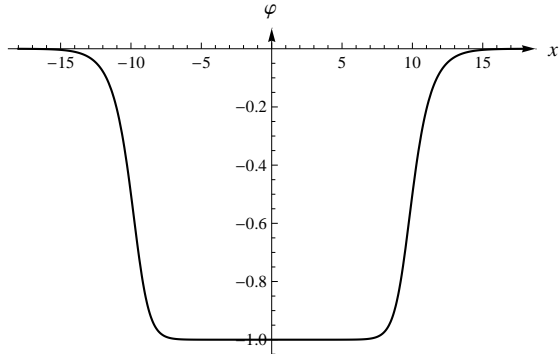}
	\caption{\label{fig:kink2} Ansatz~(\ref{eq:kink2}) at $b=10$.}
\end{figure}

\begin{figure}
	\includegraphics[scale=0.8]{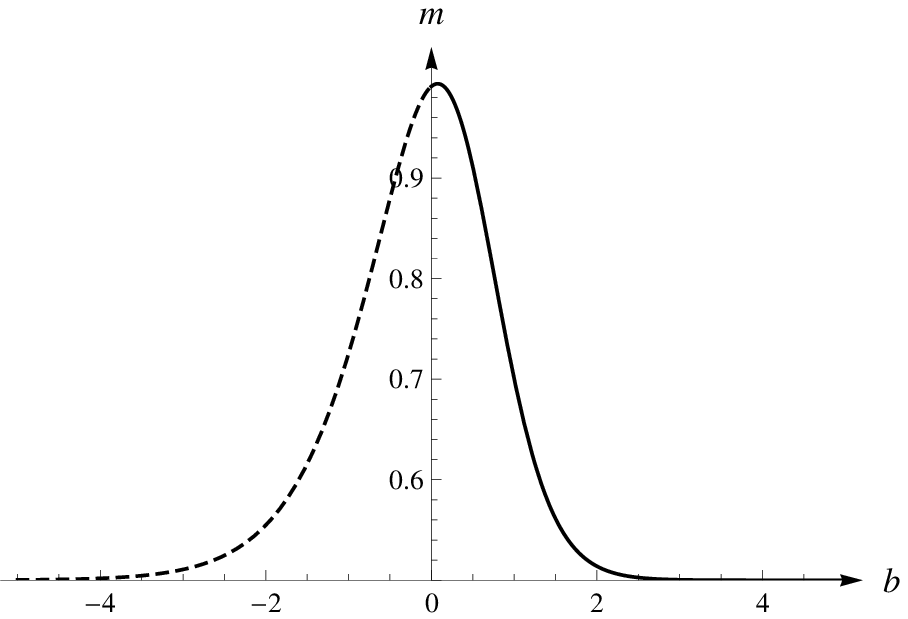}
	\caption{\label{fig:kink2-mb} The dependence $m(b)$ for the configuration~(\ref{eq:kink2}).}
\end{figure}

\begin{figure}
	\includegraphics[scale=0.8]{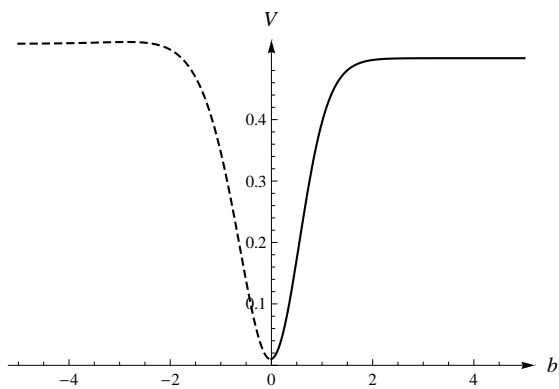}
	\caption{\label{fig:kink2-vb} The dependence $V(b)$ for the configuration~(\ref{eq:kink2}).}
\end{figure}

The initial ansatz for this type of configurations is
\begin{equation}
	\varphi_{(0,-1,0)}(x)=\varphi_{(0,-1)}(x+b)+\varphi_{(-1,0)}(x-b)+1,
	\label{eq:kink2}
\end{equation}
with $b(0)=\mbox{const}\gg 1$, $|\dot{b}(0)|=v_{\scriptsize \mbox{in}}$ defining the initial positions and velocities of the colliding kinks, see Fig.~\ref{fig:kink2}. The effective Lagrange function parameters are
\[
	m(b)=I_{+}(b),
\]
\[
	V(b)=\frac{1}{2}I_{-}(b)+\frac{1}{2}\int_{-\infty}^{\infty}\varphi_{(0,-1,0)}^2(x)(1-\varphi_{(0,-1,0)}^2(x))^2dx,
\]
where
\[
	I_{\pm}(b)=\frac{1}{2}\pm18e^{-4b}\int_{-\infty}^{\infty}\frac{dx}{(1+3e^{2(x-b)})^{3/2}(1+3e^{-2(x+b)})^{3/2}}.
\]
The plots of $m(b)$ and $V(b)$ are shown in Figs.~\ref{fig:kink2-mb} and \ref{fig:kink2-vb}.

The function $m(b)$ reaches its maximum at a small positive value of the variable $b$. In addition, the curve in Fig.~\ref{fig:kink2-mb} is not symmetric with respect to the vertical line passing through its maximum.

\begin{figure}
	\includegraphics[scale=0.8]{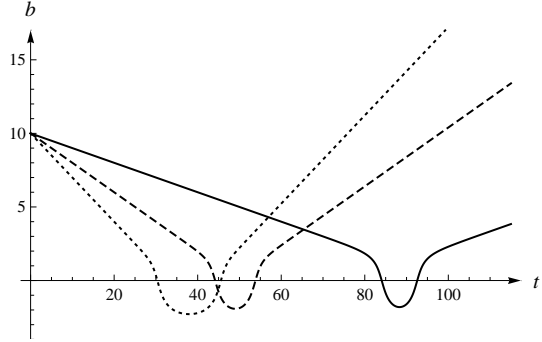}
	\caption{\label{fig:kink2-bt-f} The dependence $b(t)$ for the configuration (\ref{eq:kink2}) for $b(0)=10$ and different initial velocities below $v_{\scriptsize \mbox{cr}}^{\scriptsize \mbox{(eff)}}$: $v_{{\scriptsize \mbox{in}}}=0.1$ (solid curve), $v_{{\scriptsize \mbox{in}}}=0.2$ (dashed curve), $v_{{\scriptsize \mbox{in}}}=0.3$ (dotted curve). Collective coordinate approximation.}
\end{figure}

\begin{figure}
	\includegraphics[scale=0.8]{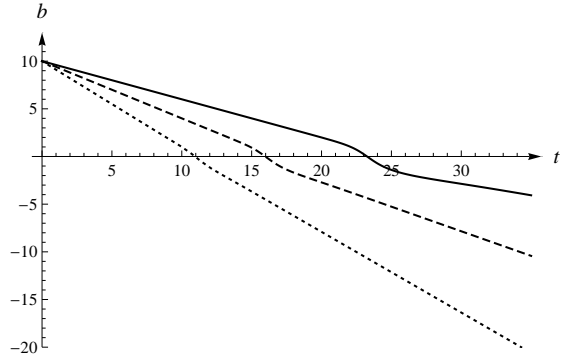}
	\caption{\label{fig:kink2-bt-g} The dependence $b(t)$ for the configuration (\ref{eq:kink2}) for $b(0)=10$ and different initial velocities above $v_{\scriptsize \mbox{cr}}^{\scriptsize \mbox{(eff)}}$: $v_{{\scriptsize \mbox{in}}}=0.4$ (solid curve), $v_{{\scriptsize \mbox{in}}}=0.6$ (dashed curve), $v_{\scriptsize \mbox{in}}=0.9$ (dotted curve). Collective coordinate approximation.}
\end{figure}

The plot of $V(b)$ has the form of an asymmetric well, and $V_{{\scriptsize \mbox{min}}}>0$ albeit close to zero. Quite importantly, $V$ tends to different asymptotical values as $b$ becomes large and positive or large and negative; numerically,
\[
	V_1=\lim_{b \to +\infty} V(b)\approx 0.500,
\]
\[
	V_2=\lim_{b \to -\infty} V(b)\approx 0.523.
\]
When $b$ is large and positive, the overlap of the two kinks in (\ref{eq:kink2}) is exponentially small; at the same time, each kink is the exact BPS-saturated solution of its topological sector. For this reason, the value of $V_1$ is simply the sum of the masses of the solitary kinks $\varphi_{(-1,0)}(x-b)$ and $\varphi_{(0,-1)}(x+b)$. The situation changes when $b$ is large and negative, which corresponds to the passage of the two kinks through each other during the collision. As the result, the two kinks change their topological sectors, analogously to the collision in the sector $(-1,0,1)$ considered in Sec.~\ref{sec:level6}: now, the kink $\varphi_{(0,-1)}(x+b)$ connects the vacua 1 and 0, whereas the kink $\varphi_{(-1,0)}(x-b)$ connects the vacua 0 and 1 [the ansatz (\ref{eq:kink2}) in this case also changes its type to $(0,1,0)$]. Again, the two kinks are not the exact BPS-saturated solutions of their new topological sectors, which translates into $V_2$ being larger than $V_1$ and, as in the previously discussed collision configuration, into the existence of two collision regimes and the critical velocity that separates these two regimes. As before, the case $v_{\scriptsize \mbox{in}}<v_{\scriptsize \mbox{cr}}^{\scriptsize \mbox{(eff)}}$ corresponds to the elastic reflection of the two kinks, whereas the values of $v_{\scriptsize \mbox{in}}>v_{\scriptsize \mbox{cr}}^{\scriptsize \mbox{(eff)}}$ result in the transition of the kinks through each other and their escape to infinity [in this latter case the final configuration is of the type $(0,1,0)$]. Numerically we find $v_{\scriptsize \mbox{cr}}^{\scriptsize \mbox{(eff)}}\approx 0.32485$. Note that the critical velocity can also be estimated from a simple classical mechanics argument, which yields $v_{\scriptsize \mbox{cr}}^{\scriptsize \mbox{(mech)}}=0.302$, in a good agreement with the collective coordinate method. In Figs.~\ref{fig:kink2-bt-f} and \ref{fig:kink2-bt-g} we show the plots of $b(t)$ obtained from (\ref{eq:eiler}) with the obvious replacement $a\to b$, for different initial velocities.

\begin{figure}
	\includegraphics[scale=0.8]{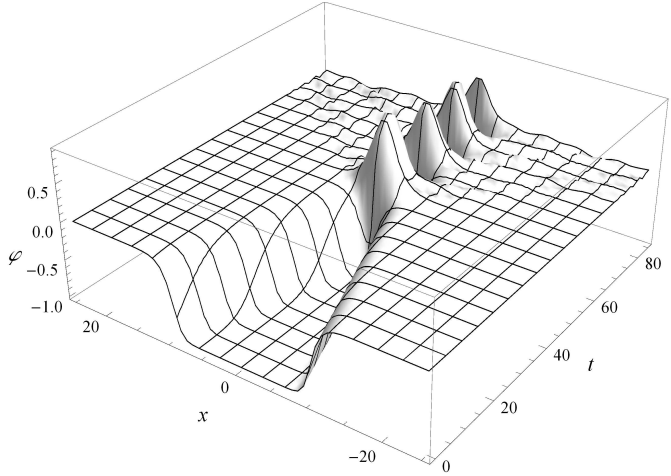}
	\caption{\label{fig:kink2-eom-f-surf} Space-time picture of the evolution of the configuration $(0,-1,0)$ for $b(0)=10$ and $v_{\scriptsize \mbox{in}}=0.2$. Numerical solution of Eq.~(\ref{eq:eom}).}
\end{figure}

\begin{figure}
	\includegraphics[scale=0.8]{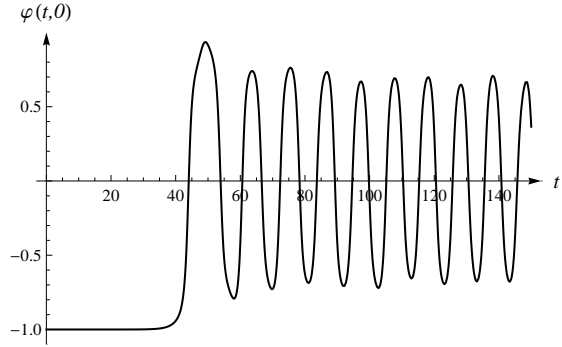}
	\caption{\label{fig:kink2-eom-f-fi0} The dependence $\varphi(t,0)$ for the configuration (\ref{eq:kink2}) for $b(0)=10$ and $v_{\scriptsize \mbox{in}}=0.2$.}
\end{figure}

As in the previous case, we also studied the evolution of the initial configuration of the type (\ref{eq:kink2}), solving the exact equation of motion (\ref{eq:eom}) numerically, with the same initial separation and initial velocities. Note that, in order to account for the relativistic effects due to the motion of the kinks properly, one can simply apply Lorentz boosts to stationary kinks. All calculations were performed at $b(0)=10$. In contrast to the case of the collective coordinate approximation, at low initial velocities we observed the formation of a quasibound state of the two kinks $(0,-1)$ and $(-1,0)$ at low initial velocities $v_{\scriptsize \mbox{in}}<v_{\scriptsize \mbox{cr}}^{\scriptsize \mbox{(eom)}} \approx 0.289$. This value of $v_{\scriptsize \mbox{cr}}^{\scriptsize \mbox{(eom)}}$ reproduces the corresponding result of Ref.~\cite{dorey01}. The evolution of this configuration in time can be described as follows. After the first collision the kinks pass through each other, forming a configuration of the type $(0,1,0)$, and continue moving until there is a certain (negative) distance between the kinks. At this point, the kinks stop and reverse, passing through each other for the second time and returning to the configuration $(0,-1,0)$. These steps are then repeated, with the maximal distance between the kinks getting smaller with each successive collision, which is illustrated in Fig.~\ref{fig:kink2-eom-f-surf}. A plot of $\varphi(t,0)$ for $v_{\scriptsize \mbox{in}}=0.2$ is shown in Fig.~\ref{fig:kink2-eom-f-fi0}.


The formation of the quasibound state of the two kinks (the so-called bion) can be qualitatively explained by a repeated radiation of small waves. Indeed, as the two initial kinks move through each other and change their topological sectors, they cease being ``almost exact'' solutions of their sectors, as discussed above. This results in the kinks starting to decay into the ``true'' kinks of their new sectors, emitting small waves and losing energy and momentum. Then, due to the mutual attraction between the kinks (see Fig.~\ref{fig:kink2-vb}), the motion of the kinks is reversed and they pass through each other again, returning to their initial topological sectors. However, due to the emission of the small waves at the preceding stage of the collision, both kinks are no longer the ``almost exact'' solutions of their initial sectors, hence they continue emitting small waves, again decaying into the true kinks of these sectors and radiating away more energy and momentum. As the result of this relaxation process, a long-lived quasibound state of the kinks is formed. This state is continuously emitting small waves and slowly decaying.

\begin{figure}
	\includegraphics[scale=0.8]{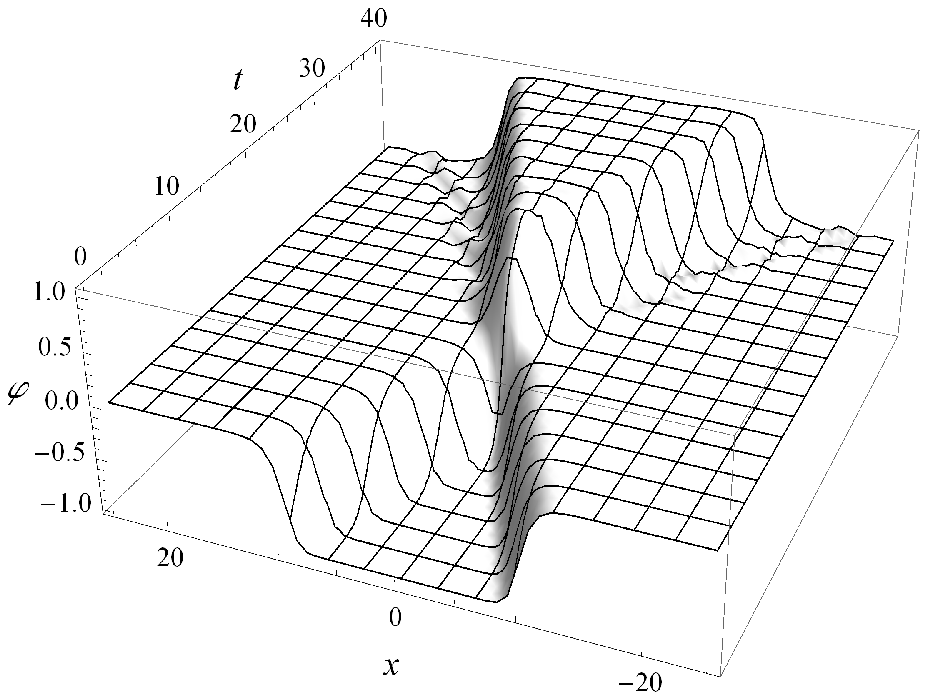}
	\caption{\label{fig:kink2-eom-g-surf} Space-time picture of the evolution of the configuration $(0,-1,0)$ for $b(0)=10$ and $v_{\scriptsize \mbox{in}}=0.6$. Numerical solution of Eq.~(\ref{eq:eom}).}
\end{figure}

\begin{figure}
	\includegraphics[scale=0.8]{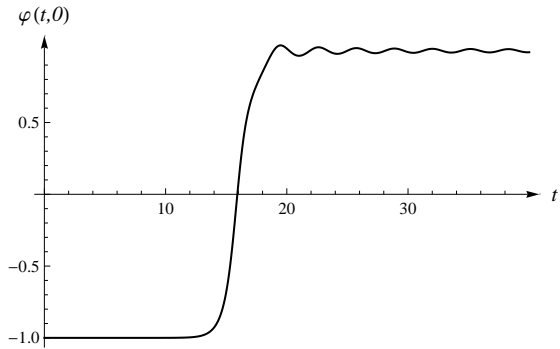}
	\caption{\label{fig:kink2-eom-g-fi0} The dependence $\varphi(t,0)$ for the configuration (\ref{eq:kink2}) for $b(0)=10$ and $v_{\scriptsize \mbox{in}}=0.6$.}
\end{figure}

On the other hand, the evolution of the initial ansatz in the regime $v_{\scriptsize \mbox{in}}>v_{\scriptsize \mbox{cr}}^{\scriptsize \mbox{(eom)}}$, as results from numerically solving the exact equations of motion, is qualitatively similar to what is obtained in the collective coordinate approximation. Namely, the kinks pass through each other, form a configuration of the type $(0,1,0)$, and escape to infinities. This is illustrated in Figs.~\ref{fig:kink2-eom-g-surf} and \ref{fig:kink2-eom-g-fi0}, where the space-time evolution and the plot of $\varphi(t,0)$ are shown for $v_{\scriptsize \mbox{in}}=0.6$.

We see therefore that the collective coordinate approximation results in either elastic reflection of the two kinks at $v_{\scriptsize \mbox{in}}<v_{\scriptsize \mbox{cr}}^{\scriptsize \mbox{(eff)}}$ or in their transition through each other at $v_{\scriptsize \mbox{in}}>v_{\scriptsize \mbox{cr}}^{\scriptsize \mbox{(eff)}}$, and no (quasi-)bound state of the two kinks can be found. The comparison with the numerical solution of the exact equations of motion shows that the collective coordinate method fails to describe the details of the low-velocity kink-kink collisions, in particular, the absence of elastic reflection at all velocities. Note that the formation of a long-lived bound state of two kinks has been known for models such as $\lambda \varphi^4$ and sine-Gordon for a long time, see, e.g., \cite{aek01} and \cite{aek02} and references therein.

Note also that the formation of the bound state can be modeled in the collective coordinate framework. This can be achieved, for instance, by introducing effective friction in the system of the two kinks, which would allow for a loss of energy and hence make the system fall into the potential well in Fig.~\ref{fig:kink2-vb} -- thus forming a bound state of the two kinks. In this connection, we refer the reader to, e.g., Ref.~\cite{aek02}.

\subsection{\label{sec:level8} Evolution of the configuration $(-1,0,-1)$}

\begin{figure}
	\includegraphics[scale=0.8]{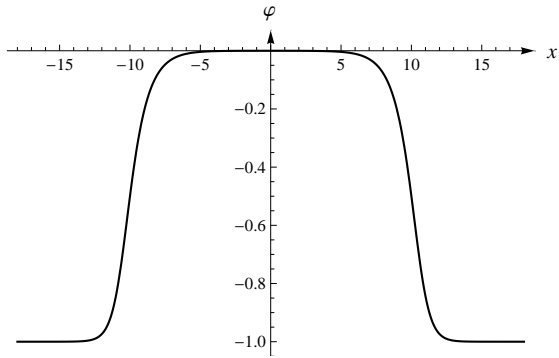}
	\caption{\label{fig:kink3} Ansatz~(\ref{eq:kink3}) at $c=10$.}
\end{figure}

\begin{figure}
	\includegraphics[scale=0.8]{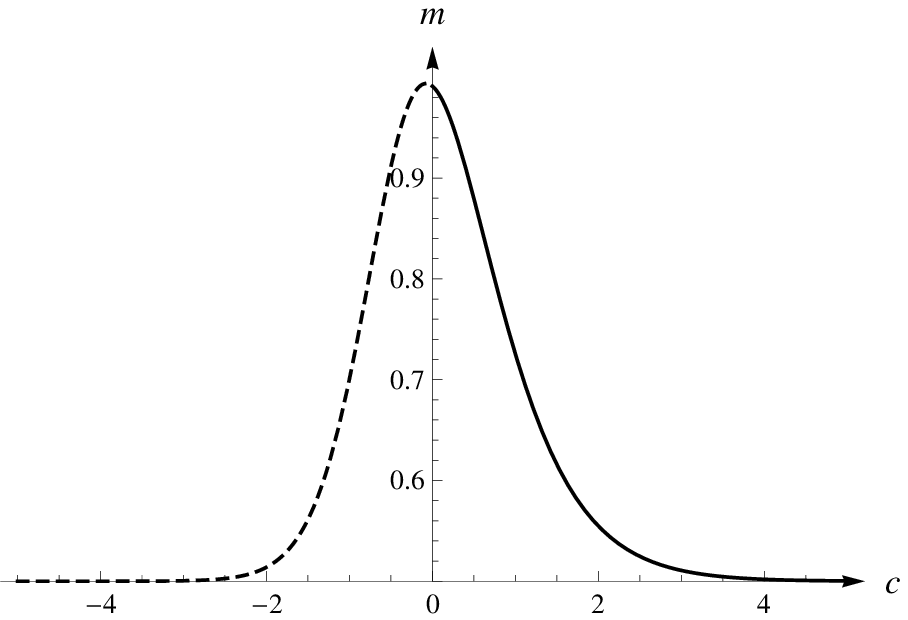}
	\caption{\label{fig:kink3-mc} The dependence $m(c)$ for the configuration~(\ref{eq:kink3}).}
\end{figure}

\begin{figure}
	\includegraphics[scale=0.8]{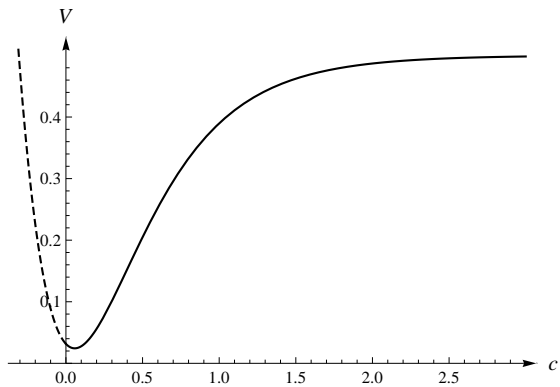}
	\caption{\label{fig:kink3-vc} The dependence $V(c)$ for the configuration~(\ref{eq:kink3}).}
\end{figure}

Finally, we consider the collision of the kinks $(-1,0)$ and $(0,-1)$. The corresponding initial configuration is
\begin{equation}
	\varphi_{(-1,0,-1)}(x)=\varphi_{(-1,0)}(x+c)+\varphi_{(0,-1)}(x-c).
	\label{eq:kink3}
\end{equation}
The plot of $\varphi_{(-1,0,-1)}(x)$ at $c=10$ is shown in Fig.~\ref{fig:kink3}. The effective Lagrange function parameters are
\[
	m(c)=I_{+}(c),
\]
\[
	V(c)=\frac{1}{2}I_{-}(c)+\frac{1}{2}\int_{-\infty}^{\infty}\varphi_{(-1,0,-1)}^2(x)(1-\varphi_{(-1,0,-1)}^2(x))^2dx,
\]
where
\[
	I_{\pm}(c)=\frac{1}{2}\pm18e^{4c}\int_{-\infty}^{\infty}\frac{dx}{(1+3e^{-2(x-c)})^{3/2}(1+3e^{2(x+c)})^{3/2}}.
\]
The plots $m(c)$ and $V(c)$ are shown in Figs.~\ref{fig:kink3-mc} and \ref{fig:kink3-vc}.

\begin{figure}
	\includegraphics[scale=0.8]{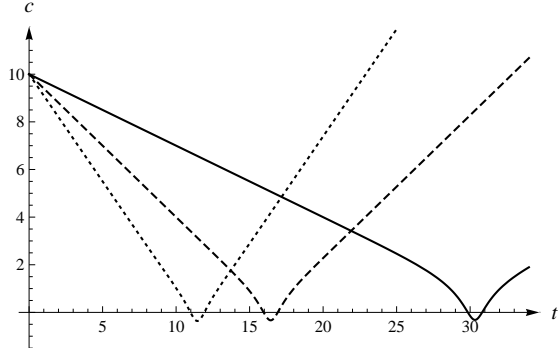}
	\caption{\label{fig:kink3-ct} The dependence $c(t)$ for the configuration (\ref{eq:kink3}) for $c(0)=10$ and different initial velocities: $v_{{\scriptsize \mbox{in}}}=0.3$ (solid curve), $v_{{\scriptsize \mbox{in}}}=0.6$ (dashed curve), $v_{{\scriptsize \mbox{in}}}=0.9$ (dotted curve). Collective coordinate approximation.}
\end{figure}

Note that at $c<0$ the ansatz (\ref{eq:kink3}) becomes a configuration of the type $(-1,-2,-1)$. Hence the potential $V(c)$ increases linearly with $|c|$, due to $\varphi=-2$ not being a vacuum of the $\varphi^6$ model. As a consequence, the colliding kinks can penetrate each other only by small distance and we expect to observe an elastic reflection of the two kinks at any initial velocity. To illustrate that, we show in Fig.~\ref{fig:kink3-ct} the profiles of $c(t)$ for several values of the initial velocity $|\dot{c}(0)|$.

\begin{figure}
	\includegraphics[scale=0.8]{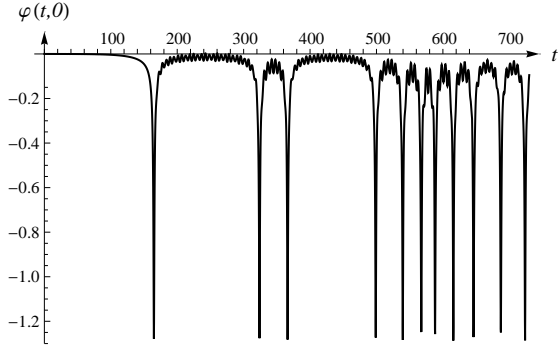}
	\caption{\label{fig:kink3-eom-bion-fi0} The dependence $\varphi(t,0)$ for the configuration (\ref{eq:kink3}) for $c(0)=10$ and $v_{\scriptsize \mbox{in}}=0.043$. Bion formation.}
\end{figure}

\begin{figure}
	\includegraphics[scale=0.8]{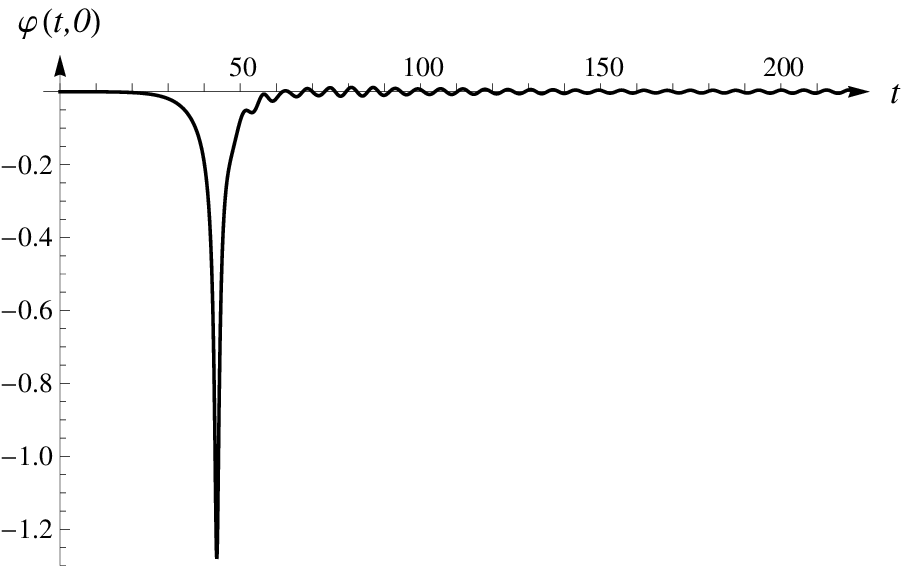}
	\caption{\label{fig:kink3-eom-g-fi0} The dependence $\varphi(t,0)$ for the configuration (\ref{eq:kink3}) for $c(0)=10$ and $v_{\scriptsize \mbox{in}}=0.2$.}
\end{figure}

The numerical study of the exact evolution in this sector shows that, similarly to the type $(0,-1,0)$ considered before, there is a critical velocity $v_{\scriptsize \mbox{cr}}^{\scriptsize \mbox{(eom)}}\approx 0.0448$ such that at $v_{\scriptsize \mbox{in}}<v_{\scriptsize \mbox{cr}}^{\scriptsize \mbox{(eom)}}$, a slowly decaying quasibound state of the two kinks is formed (with an important exception discussed below), whereas at $v_{\scriptsize \mbox{in}}>v_{\scriptsize \mbox{cr}}^{\scriptsize \mbox{(eom)}}$ the two kinks are reflected off each other almost elastically. These collision regimes are illustrated by Figs.~\ref{fig:kink3-eom-bion-fi0} and \ref{fig:kink3-eom-g-fi0}, in order. Note that the value of $v_{\scriptsize \mbox{cr}}^{\scriptsize \mbox{(eom)}}=0.0448$ that we obtain differs slightly from that of Ref.~\cite{dorey01} that quotes $v_{\scriptsize \mbox{cr}}=0.0457$.

\begin{figure}
	\includegraphics[scale=0.8]{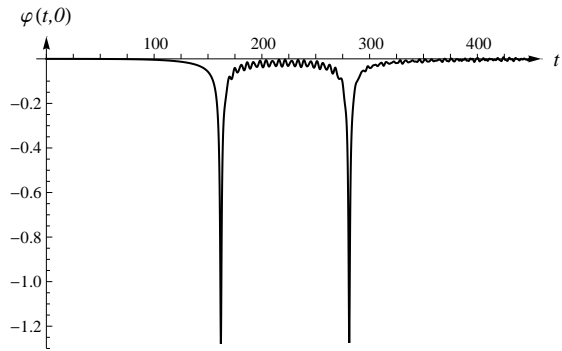}
	\caption{\label{fig:kink3-eom-okno-fi0} The dependence $\varphi(t,0)$ for the configuration (\ref{eq:kink3}) for $c(0)=10$ and $v_{\scriptsize \mbox{in}}=0.044$. Escape window.}
\end{figure}

A very peculiar feature of the collisions in this sector (also not captured by the collective approximation) is the so-called escape windows, narrow ranges of the initial velocity in the domain $v_{\scriptsize \mbox{in}}<v_{\scriptsize \mbox{cr}}^{\scriptsize \mbox{(eom)}}$ where the kinks escape to infinity after two (three, etc.) collisions instead of forming the bion, Fig.~\ref{fig:kink3-eom-okno-fi0}. Our numerical calculation finds an escape window at $v_{\scriptsize \mbox{in}}\approx 0.04420$ and a few more escape windows at different values of $v_{\scriptsize \mbox{in}}$ in the range $\left[0.04423;0.04428\right]$, whereas at $v_{\scriptsize \mbox{in}}<v_{\scriptsize \mbox{cr}}^{\scriptsize \mbox{(eom)}}$ outside the escape windows we observe the formation of a long-lived bion.

The origin of the escape windows is the resonant energy exchange: during the first collision, a part of the kink's kinetic energy is transferred into the vibrational mode of the kink, which cannot then escape to infinity after the first reflection, hence the kinks collide again. If a certain resonance condition holds, energy stored in the vibrational mode can be returned back into the kinetic energy, and the kinks are then able to escape. Originally, escape windows were discovered in the $\lambda\varphi^4$ model~\cite{campbell01} and in a modified sine-Gordon model~\cite{campbell02}. Note that the resonant energy exchange can also occur not in the second but in any subsequent collision, so that the kinks escape to infinities after colliding three, four times, etc.

In contrast to the $\lambda\varphi^4$ model, the $\varphi^6$ model does not have a vibrational (shape) mode in the kinks' excitation spectra. Instead, in the collisions of $\varphi^6$ kinks a vibrational mode of the two kinks is excited, as shown in Ref.~\cite{dorey01} (where a very comprehensive study of escape windows in the $\varphi^6$ model was given). A similar mechanism is involved in the so-called quasiresonances in the double sine-Gordon model \cite{gani06}.

We solved the partial differential equation (\ref{eq:eom}) and the ordinary differential equation (\ref{eq:eiler}) with the use of the computer algebra system ``Mathematica''. The ordinary differential equation was solved by the Livermore Solver for Ordinary Differential Equations with the automatic switching for nonstiff (Adams) and stiff (BDF) methods). The partial differential equation was solved by the method of lines in the domain $x\in\left[-l,l\right]$ and $t\in\left[0,T\right]$ [$l\gg a(0)$, $b(0)$, or $c(0)$], with the appropriate initial and boundary conditions for each initial kink-kink configuration.

\section{\label{sec:level9} Conclusion}

The aim of the present study has been a detailed comprehensive investigation of the applicability of the collective coordinate approximation with $1$ degree of freedom to modeling of the kink-kink collision processes. Unlike previous authors, we performed modeling in all the configuration types [note that in the pairs $(0,-1,0)$ and $(0,1,0)$, $(-1,0,-1)$ and $(1,0,1)$, as well as $(-1,0,1)$ and $(1,0,-1)$, each of the two configurations is dynamically equivalent to the other; hence, only one configuration is needed to be considered in each pair]. We also compare the results of both methods and draw conclusions about the applicability of the collective coordinate approximation in each of these cases. In the sector $(-1,0,1)$ we do not confirm the discrepancy reported by the author of Ref.~\cite{goatham}. In the sectors $(0,-1,0)$ and $(-1,0,-1)$ we discovered, solving the exact equation of motion (\ref{eq:eom}) numerically, that the two colliding kinks can form a long-lived bound state (a bion). We provide a qualitative explanation of the process of the bion formation, based on the dynamics of a kink that leaves its topological sector and stops being a nearly exact solution. In the sector $(-1,0,-1)$, we confirm the existence of escape windows.

The kink scattering in the sector $(-1,0,1)$ at sufficiently small initial velocities can be modeled reasonably well in the framework of the collective coordinate approach. The basis for this conclusion is a good agreement between the results of the two methods used in this work.

At the same time, the exact evolution of the kinks in low-velocity $v_{\scriptsize \mbox{in}}<v_{\scriptsize \mbox{cr}}^{\scriptsize \mbox{(eom)}}$ collisions exhibits a capture of the kinks by each other and the resulting formation of a long-lived quasibound state of the two kinks. This phenomenon occurs in the kink-kink collisions where the asymptotics at the positive and the negative infinities are the same, i.e.,\ in the sectors $(0,-1,0)$ and $(-1,0,-1)$. Note that in all these cases the interaction between the kinks is attractive. However, within the framework of the collective coordinate approximation without effective friction it appears impossible to describe a bound state of the two kinks.

In the sector $(0,-1,0)$ the kinks can pass through each other and escape to infinity. This process is observed both in the collective coordinate approach and in the exact numerical dynamics. The values of the critical velocity obtained by the two methods are rather close (within $10\%$).

In addition, one encounters the resonant energy exchange between the translational and the vibrational modes of the two colliding kinks in the sector $(-1,0,-1)$. This mechanism leads to the escape windows at $v_{\scriptsize \mbox{in}}<v_{\scriptsize \mbox{cr}}^{\scriptsize \mbox{(eom)}}$, which are well studied and described for the $\varphi^6$ model in \cite{dorey01}. Of course, within the collective coordinate approximation with $1$ degree of freedom we can observe no resonance phenomena. Note that in Ref.~\cite{belova01} it was shown, however, that the escape windows can be reproduced within the collective coordinate method if the interaction with the vibrational mode of the kink is taken into account. A recent publication \cite{weigel} should be mentioned here, where the vibrational modes of the kinks have been taken into account in the framework of the collective coordinate approach. It allows one to simulate resonance phenomena such as the escape windows or the quasiresonances.

Thus, the collective coordinate approximation with $1$ degree of freedom provides, in general, a good description of the kink-kink collision process at intermediate initial velocities. At low initial velocities in the sectors $(0,-1,0)$ and $(-1,0,-1)$ the exact dynamics is qualitatively different from that predicted by the collective coordinate approximation. At the same time, at ultrarelativistic initial velocities notable quantitative discrepancies are observed.

Note that in the sector $(-1,0,1)$ the interaction between the kinks has a repulsive character and a bound state of the kinks cannot be formed. The collective coordinate approach works well also for small initial velocities in that sector.

Our analysis of the collective coordinate approximation applied to kink-kink collisions in the $\varphi^6$ model demonstrates the following apparently rather general properties of this method. First, the applicability of the collective coordinate approximation is limited to nonrelativistic kinks' velocities, since the chosen ansatz and the effective Lagrange function are not Lorentz invariant by construction. This results in large deviations from the exact evolution at relativistic velocities of colliding kinks. Although this shortcoming of the method can be fixed by properly taking into account the Lorentz invariance, this would also result in significantly more complex equations of motion describing the dynamics of the collective coordinate. Second, there are phenomena occurring at low velocities of the colliding kinks in the presence of an effective attractive interaction between the kinks, such as the resonant energy exchange or the kink-kink capture. The collective coordinate approximation with a single collective coordinate does not include the degrees of freedom that are responsible for the correct description of these phenomena. For instance, this method does not reproduce the escape windows or the formation of a bion. On the other hand, the regime when the two kinks only interact with each other for a short time (e.g., when there is a repulsive force between the kinks or when their collision velocity is large enough) is described well within the collective coordinate approximation. Apparently, these limitations of the method are quite general and hence our analysis of the $\varphi^6$ kink-kink collisions could provide a ``rule of thumb'' useful in studies of other models in the collective coordinate approximation.

In conclusion, we emphasize that the present study opens wide prospects for further research. In particular, as already mentioned, the kink $(-1,0)$ in the sector $(0,1)$ starts to decay and as a consequence changes its velocity. We suppose that this phenomenon may be described by analyzing the impact of the decay on the kink's zero (translational) mode.

\section*{ACKNOWLEDGMENTS}

The authors are very grateful to Dr.~V.~Lensky for many fruitful discussions 
and numerous critical comments that resulted in great improvement of the 
manuscript. This work was supported by the Russian Federation Government under Grant No.~NSh-3830.2014.2. V.~A.~Gani also acknowledges the support of the Young Faculty Member Program of the National Research Nuclear University MEPhI.

\end{document}